\documentclass[preprint,showpacs,preprintnumbers,amsmath,amssymb]{revtex4}

\usepackage{graphicx}
\usepackage{dcolumn}
\usepackage{bm}

\begin{document}

\title{Primoridal black hole minimum mass}
\author{James R. Chisholm}
\affiliation{Institute for Fundamental Theory, University of Florida, Gainesville, Florida 32611-8440, USA}

\date{\today}

\begin{abstract}

In this paper we revisit thermodynamic constraints on primordial black hole (PBH) formation in the early universe.  Under
the assumption that PBH mass is equal to the cosmological horizon mass, one can use the 2nd Law of Thermodynamics to put
a lower limit on the PBH mass.  In models of PBH formation, however, PBHs are created at some fraction of the horizon mass.
We show that this thermodynamic constraint still holds for sub-horizon PBH formation.

\end{abstract}


\maketitle

\section{Introduction}

The primary utility of primordial black holes (PBHs) \citep{zeldovich, hawking1} in cosmology today is their absence; their lack of observation 
allows one to put a limit on the size of density perturbations in the early universe \citep{kim1,GreLid1}. 
Light ($M_{BH}< 10^{15}$g) PBHs will evaporate by the present day due to Hawking radiation emission \citep{hawking2}.  The injection
of that radiation into the universe can have observable consequences \citep{okele,lindley1,macgibbon3,halzen,lemoine,barrau1,barrau2,khlopov2} unless the PBH 
number density is sufficiently small.  The PBH lifetime is related to the PBH mass, and the PBH mass is proportional
to the horizon mass at which it formed.  We therefore can use the lack of PBH emission at a certain time (the time of Big Bang nucleosynthesis until the present day) to
constrain PBH formation at a much earlier time (from matter-radiation equality back to the time of reheating).  

Of the steps involved in this, the 
most uncertain is the relation between the PBH mass ($M_{BH}$) and the horizon mass ($M_H$) at which they formed.
The mass of a primordial black hole formed at a time $t$ is
\begin{equation}\label{mbh}
M_{BH} = f M_H,
\end{equation}
where the horizon mass $M_H$ is defined through
\begin{equation}\label{mh}
M_H = \frac{4}{3}\pi \rho R_H^3
\end{equation}
with the horizon size $R_H = H^{-1}$ and the Hubble parameter $H$ given by the Friedmann equation
\begin{equation}\label{friedmann}
H^2 = \frac{8\pi}{3} G\rho.
\end{equation}
The function $f$ is the fraction of the horizon mass that goes into the PBH.    An analytic calculation
in \citet{carr3} gave $f = w^{3/2}$, where $w = p/\rho$ is the background equation of state.  For a radiation dominated
universe, $w = 1/3$ and so $f \approx 0.2$, close enough so that most authors simply assumed $f \sim 1$.  The upper bound 
on $f$ for a radiation dominated universe is unity, though it is lower for backgrounds with stiffer equations of 
state \cite{harada}.

The formation of a black hole changes the entropy of the universe.  In the early universe, the formation of PBHs leads
to isocurvature (entropy) as well as adiabatic (density) perturbations \citep{chisholm}.  The thermodynamic constraint on PBH mass,
first considered in \citet{lee}, derives from the 2nd Law of Thermodynamics: that
the entropy of a closed system can only increase.  There, the horizon-sized region of the universe where PBH formation is 
happening is the closed system (in that it is 
initially causally disconnected from the rest of the universe).  The entropy change we consider is that between 
the initial state radiation and the final state PBH-radiation system.  For PBH creation exactly at the horizon mass ($f=1$),
there is no final state radiation and it is straightforward to compute the difference in entropy.  The initial state 
consists of a perturbed FRW universe filled with a perfect radiation fluid of known density (and thus known horizon mass).
The final state consists of a PBH equal to the horizon mass with temperature and entropy given by
\begin{equation}
T_{BH} = \frac{\kappa}{2 \pi} = \frac{1}{8\pi G M_{BH}},
\end{equation}
\begin{equation}\label{spbh}
S_{BH} = \frac{A}{4G} = 4\pi G M_{BH}^2,
\end{equation}
where $A$ and $\kappa$ are the PBH surface area and surface gravity, respectively.  That the PBH must have more entropy
than the initial radiation can be turned into the limit $M_{PBH} \gtrsim 0.37 M_P$ \citep{lee}\footnote{We use a different
definition of the horizon radius than \citet{lee}, which accounts for the discrepancy between their Equation~20 and our result.}.
We go through the derivation of this limit in a later section.  Note, however, that the Lee bound is coincident with
the implicit quantum ``bound'' at the Planck mass where quantum gravity becomes important.  There we are no longer certain
about the physics of black hole creation.  For a horizon sized PBH, the thermodynamic bound is not higher than this 
quantum bound.  It was not known is this continues to be true for sub-horizon PBH formation, which is the more physical
situation.  We prove in this paper that it does.

\section{Sub-horizon PBH formation}

It was observed in \cite{choptuik,evans} that gravitational collapse near threshold exhibits self-similar scaling behavior.  
The consequence of this for PBHs is that $f$ is not a constant as shown above but rather depends on the size of the perturbation.
For a PBH formed from an 
overdensity $\delta = (\rho -\bar \rho)/\bar\rho$ being above a threshold $\delta_C$, the fraction is given by 
the power law
\begin{equation}
f = \kappa (\delta - \delta_C)^\gamma,
\end{equation}
where $\kappa \approx 3$, $\gamma \approx 0.37$ and $\delta_C \approx 2/3$ for PBH formation during the
radiation dominated era.  Formally $f$ vanishes as $\delta \rightarrow \delta_C$.  

Numerical simulations of gravitational collapse \citep{niemeyer,hawke,musco} verify this power-law behavior, though
the values of the parameters ($\gamma, \kappa, \delta_C$) differ slightly for different shapes of the initial density
perturbation.  The profiles most common in numerical studies are a Gaussian, a Mexican hat and a sixth order polynomial 
(see Figure~1 of \cite{niemeyer} for definitions and shapes), and we list the parameters in Table~\ref{table}.  While the
values of $\kappa$ and $\delta_C$ vary by more than a factor of 2, the slope $\gamma$ is within 1\% of the value  
$\gamma = 0.3558...$ computed analytically in \citet{koike}.

\begin{table}
\caption{\label{table}  Collapse parameters for different density perturbation profiles. }
\begin{ruledtabular}
\begin{tabular}{lccc}
Reference & Gaussian & Mexican hat & Polynomial\\
\hline
\citet{niemeyer} & $\gamma$ = 0.34 & $\gamma$ = 0.36 & $\gamma$ = 0.37\\
                               & $\delta_C$ = 0.70 & $\delta_C$ = 0.67 & $\delta_C$ = 0.71\\
                               & $\kappa$ = 11.9 & $\kappa$ = 2.9 & $\kappa$ = 2.4\\
\hline
\citet{hawke} & $\gamma$ = 0.35 & - & - \\
                          &  $\delta_C$ = 0.4 & - & - \\
\hline
\citet{musco} & \multicolumn{3}{c}{$\gamma \approx$ 0.36 - 0.37}\\
                          & $\delta_C$ = 0.71 & $\delta_C$ = 0.67 & $\delta_C$ = 0.71\\
\end{tabular}
\end{ruledtabular}
\end{table}

The results of \cite{hawke} further suggest that $f$ reaches a minimum value at $\sim 10^{-3.5}$.  Exactly how and why
this occurs is not yet known, as this result has not yet been seen in other numerical simulations.  If further studies
show this bound to be a numerical artifact, then it remains that a PBH may have an arbitrarily small mass compared to the
horizon at which it formed.  We show in this short paper that there is indeed a lower bound on $f$ coming from entropy
constraints.

\section{Entropy constraint}

For subhorizon PBH formation ($f < 1$), one must take into account the entropy of the radiation outside of the event 
horizon but within the cosmological
horizon.  Using entropy to constrain PBH formation then becomes more subtle as the PBH is now no longer an isolated region.
The entropy of the final state radiation is initial condition dependent ({\it e.g.}, the shape of the perturbation as 
discussed in the previous section) and will have contributions from shocks.  

To be as general as possible, we do not attempt to account for the total amount of entropy in the horizon volume during
collapse.  Instead, we consider the 
region where a PBH is formed at the exact moment of PBH creation.  During the infinitesimal time spanning the moment where
a high radiation density region formed an event horizon and thus a black hole, there is a finite change in the total 
entropy.  Any
change in entropy due to an outgoing finite radiation flux, on the other hand, will be infinitesimally small.

Our bound comes from comparing the entropy of the region immediately before and after PBH creation.  We begin with the
final state, where the PBH has a mass and entropy as given in Equation~(\ref{spbh}).  Our region of interest is then the 
sphere of radius $R_{BH} = 2 G M_{BH}$ and volume $V_{BH}$ = $4\pi R_{BH}^3/3$.  We assume throughout that the radiation 
is a perfect fluid, so that its density profile completely determines its behavior.  We can define an average density of
the radiation before PBH formation,
\begin{equation}
\rho_* = \frac{M_{BH}}{V_{BH}} = \frac{3}{21\pi}\frac{M_P^6}{M_{BH}^2} = \frac{\rho}{f^2}.
\end{equation}
where the last equality is obtained from using Equations~(\ref{mbh}~-~\ref{friedmann}).  

The entropy density for a perfect radiation fluid of density $\rho$, pressure $p$ and temperature $T$ is given by
\begin{equation}
s_{rad} = \frac{\rho + p}{T} = \frac{4}{3} \frac{\rho}{T} = \frac{4}{3} \left(\frac{g_* \pi^2}{30}\right)^{1/4} \rho^{3/4},
\end{equation}
where we have used the density temperature relation for a perfect radiation fluid with $g_*$ effective relativistic degrees of 
freedom \footnote{Since $\rho_{rad} \propto T^4$ and $s_{rad} \propto T^3$ , the temperature weighted average used to compute the effective relativistic degrees of freedom gives slightly different values, commonly labelled $g_*$ and $g_{*S}$ respectively.    We are assuming $g_* \approx g_{*S}$, which is valid for high $T$.},
\begin{equation}
\rho = \frac{\pi^2}{30}g_* T^4.
\end{equation}

The total entropy inside the region of volume $V_{BH}$ is approximated by
\begin{equation}\label{srad}
S_{rad} = \frac{4}{3} \left(\frac{g_* \pi^2}{30}\right)^{1/4} \rho_*^{3/4}V_{BH}.
\end{equation}
The condition that $S_{BH} \geq S_{rad}$ is
\begin{equation}\label{bound1}
\frac{M_{BH}}{M_P} \geq \frac{4}{27} \sqrt{\frac{g_*}{5\pi}} \approx 0.39 \sqrt{\frac{g_*}{106.75}},
\end{equation}
Or equivalently a bound on $f$:
\begin{equation}\label{bound2}
f \geq \frac{16\pi}{405} g_* \left(\frac{T}{M_P}\right)^2.
\end{equation}

This result verifes that of \citet{lee} on sub-horizon scales: that the Planck mass remains the thermodynamic lower bound on PBH mass.
Pushing this bound higher requires changing $g_*$, which is determined by the model of particle physics one uses.
Including only Standard Model
 particles, $g_* = 106.75$ at $T > 300$ GeV \cite{kolbturner}.  This will roughly double in SUSY models, though
since the bound $\propto \sqrt{g_*}$, this has only a small effect.  Only in a Hagedorn-type model \cite{hagedorn} where
the number of states increases exponentially with temperature could the bound increase appreciably.  However, in that case the
background cosmology we have assumed is likely to be invalid.  The same is true for models of extra dimensions, where
the fundamental scale of gravity is different from $M_P$.  The formation and evaporation of PBHs is different \cite{guedens}
in that case and we do not consider it further here.

\section{Conclusions}

We note that this bound might not be saturated in any astrophysical situation of PBH creation.  
This is because we have assumed throughout that PBHs form only at a single time (and thus single horizon mass).  
As PBHs will form for a range of horizon masses, the mass at which a PBH is formed (at a particular point in space)
is given by maximizing the expression for $M_{BH}$ over both $\delta$ and $M_H$.  Using an excursion set formalism, 
\citet{GreLid2} showed that PBHs still form at order the horizon mass for the both the case of a blue perturbation spectrum
and a spike at a specific scale (assuming a number density that saturates the current constraint).  It is possible, however, that
the result breaks down (so that we have $f \ll 1$) for a different type of spectrum and/or for a smaller number density of PBHs
produced.  However, in the limit that the PBH number density becomes astrophysically interesting (approaches the current
upper bounds), there is still a tight correlation between PBH mass and horizon mass \citep{yokoyama}.
Our bound, therefore, would be satisfied by a large margin for any theoretically observable PBH population as a whole, but could
be by a smaller margin for any individual PBH.

\begin{acknowledgments}
The author would like to thank Daniel Chung and Jim Fry for useful discussions.
\end{acknowledgments}
\appendix
\section{Self-similar gravitational collapse and curved space effects}
To simplify the analysis in this paper we have assumed flat (Minkowski) space and uniform radiation density, neither 
of which are strictly accurate in PBH formation.  It might, however, be possible to exploit the self-similar nature of 
the collapse to obtain a more accurate calcuation of the initial state radiation.  
Self-similar perfect fluid solutions have been extensively studied \citep{coley}.
The entropy calcuation including both curved space and using a proper density profile would be 
straight-forward though tedious.  A self-similar solution asymptotic to the FRW universe (which is our initial condition)
maintains a flat density profile at small radii, which is verified in numerical simulations \citep{hawke}.
The drawback to this method, as first noted by \citet{evans}, is that the collapsing
fluid only maintains the self-similar solution for $r << r_{BH}$ and breaks down before the PBH forms.  Thus one still
needs to consider both the initial perturbation shape and handle the formation of shocks.  The contribution of these 
effects to the (initial state) entropy is likely to be small since the radiation density (and thus the entropy density)
falls off at larger radii where self-similarity breaks down.

For similar reasons including curved space effects will not induce an entropy much larger than computed here.  For the 
metric 
\begin{equation}
ds^2 = -\alpha(r,t)^2 dt^2 + A(r,t)^2 dr^2 + r^2 d\Omega^2,
\end{equation}
our expression for the ``mass'' of the radation (physical energy density integrated over the PBH volume) is
\begin{equation}\label{appendix_mrad}
M_{rad} = 4\pi \int_0^{r_{BH}} \rho(r) A(r,t) r^2 dr.
\end{equation}
This value is finite when computed for times before event horizon formation.  At an infinitesimal amount of time before
PBH formation, $M_{rad} = M_{PBH}$.  The expression for the entropy is
\begin{equation}
S_{rad} = 4\pi \times \frac{4}{3}\left(\frac{g_* \pi^2}{30}\right)^{1/4} \int_0^{r_{BH}} \rho(r)^{3/4} A(r,t) r^2 dr.
\end{equation}
Compare these expressions to Equations~\ref{mbh},\ref{mh},\ref{srad} above.  Because the entropy scales with density
as a power less than unity, non-uniform density profiles that maintain the normalization of Equation~\ref{appendix_mrad}
will tend to decrease $S_{rad}$.  

\bibliography{entropy}

\end{document}